\documentclass[prc,superscriptaddress,showpacs,nofootinbib,
tightenlines]{revtex4}

\usepackage{epsfig}
\usepackage{slashed}
\usepackage{amssymb}
\usepackage{color}

\begin{document}

\title{New fixed points of the renormalisation group for two-body scattering}
\author{M.~C.~Birse}
\affiliation{Theoretical Physics Division, School of Physics and Astronomy,
The University of Manchester, Manchester, M13 9PL, UK}
\author{E.~Epelbaum}
\affiliation
{Institut f\" ur Theoretische Physik II, Fakult\" at f\" ur Physik und Astronomie,
Ruhr-Universit\" at Bochum, D-44780 Bochum, Germany}
\author{J.~Gegelia}
\affiliation{Institute for Advanced Simulation, Institut f\"ur Kernphysik
   and J\"ulich Center for Hadron Physics, Forschungszentrum J\"ulich, D-52425 J\"ulich,
Germany}
  \affiliation{Tbilisi State  University,  0186 Tbilisi,
 Georgia}
\date{October 7, 2015}
\begin{abstract}
We outline a separable matrix \textit{ansatz} for the potentials in
effective field theories of nonrelativistic two-body systems with short-range
interactions. We use this ansatz to construct new fixed points of the
renormalisation-group equation for these potentials.
New fixed points indicate a much
richer structure than previously recognized in the RG flows of simple
short-range potentials.

\end{abstract}

\maketitle

\section{Introduction}

The renormalisation group (RG) has proved to be a powerful tool for elucidating
the scale dependences of systems in many areas of physics \cite{Wilson:1973jj}.
(For more recent reviews, see Refs.~\cite{Morris:1998da,Meurice11}.) These
systems include ones consisting of two or three particles at low enough energies
that the motion can be treated as nonrelativistic.
In the case of two-body scattering by short-range forces, the existence of a
nontrivial fixed point of the RG was first noted by Weinberg \cite{Weinberg:um},
although he did not go on explore the RG flow in its vicinity. This idea was
further developed in Refs.~\cite{Gegelia:gn,Kaplan:1998tg,vanKolck:1998bw},
and a complete RG analysis of the flow around this fixed point was carried out
in Ref.~\cite{Birse:1998dk}.

This fixed point describes a system in the ``unitary limit", where the scattering
length is infinite. A system with a large scattering length compared to the scales
of the underlying physics can be described in terms of perturbations
around this point. The resulting expression for the scattering
amplitude is just given by the effective-range expansion \cite{Bethe:1949yr}.
This provides
a systematic organizing scheme, or ``power counting" for an effective field
theory (EFT) that has been applied to nucleon-nucleon scattering and to ultracold
atoms in traps. A review of the RG approach in nuclear physics can be found in
Ref.~\cite{Birse:2010fj}.

The RG for two-body scattering is expected to have other fixed points. For
example, there is a trivial one, corresponding to weakly interacting systems
where the scattering can be treated perturbatively \cite{Birse:1998dk}.
The existence of further, nontrivial fixed points has been conjectured \cite{cohen}
and hints of them were seen in a functional RG analysis by Harada and Kubo
\cite{Harada:2006cw}, but explicit forms for them were not found.

Here we present a systematic method for constructing an infinite number of possible
fixed points of the RG for two-body scattering by short-range forces.
For selected examples, we study the flows close to them, which determine the power counting rules for
EFTs expanded around these points. We
also construct some of the renormalised trajectories that flow from one fixed point to
another. Each of the new fixed points has at least two unstable directions and so two
or more parameters would need to be ``fine tuned'' for a physical system to be
described by it.

\section{RG flow}

Following the approach of Ref.~\cite{Birse:1998dk}, a convenient starting point
is the Lippmann-Schwinger integral equation for the $K$ matrix for
$S$-wave two-body scattering,
\begin{equation}
K(k',k,p) = V(k',k,p,\Lambda) + 2 M\, \mathcal{P}\!
\int \frac{d^3l \,\theta(\Lambda-l)}
{(2\pi)^3}\,\frac{V(k',l,p,\Lambda)\,K(l,k,p)}{p^2-l^2}\, ,
\label{LSEquationS}
\end{equation}
where $\mathcal{P}$ stands for the principal value, $M$ is the reduced mass, and
$p=\sqrt{2ME}$ is the on-shell relative momentum. The integral over the momentum
$l$ of the intermediate state has been regulated by cutting it off at $l=\Lambda$.
The solution to Eq.~(\ref{LSEquationS}) is the fully off-shell
$K$ matrix, whose matrix elements depend on the initial and final off-shell
momenta, $k$ and $k'$.
On-shell observables can be obtained from it by setting $k=k'=p$. For example,
the on-shell $K$-matrix, ${\cal K}(p)=K(p,p,p)$, is related to the phase shift
$\delta(p)$, and hence to the effective-range expansion, by
\begin{equation}
\frac{1}{{\cal K}(p)}=-\,\frac{M}{2\pi}\,p\cot\delta(p)
=-\,\frac{M}{2\pi}\left(-\frac{1}{a}+\frac{1}{2}\,r_e\,p^2+v_2\,p^4+\cdots\right).
\label{effrexp}
\end{equation}

To obtain the RG equation, we first demand that the solution to Eq.~(\ref{LSEquationS})
be independent of the cutoff $\Lambda$.
From the physical point of view it is not necessary to have a cutoff independent off-shell $K$-matrix. However, it is always possible to find an on-shell equivalent modification of the potential such that the off-shell $K$-matrix is cutoff independent.
The obtained RG equation
contains only the potential, in contrast to the potential
$V_{\mbox{\scriptsize low-}k}$ of Bogner \textit{et al.} \cite{Bogner:2003wn},
whose evolution equation involves the scattering matrix as well.
The resulting potential
$V(k',k,p,\Lambda)$ then has a well-defined evolution with $\Lambda$
\cite{Birse:1998dk},
\begin{equation}\frac{\partial V}{\partial\Lambda}
=\frac{M}{\pi^2}\,V(k',\Lambda,p,\Lambda)\,\frac{\Lambda^2}{\Lambda^2-p^2}\,
V(\Lambda,k,p,\Lambda).
\label{RGE1}
\end{equation}

Next, we express all low-energy scales in units of $\Lambda$,
$\hat p=p/\Lambda$ \textit{etc}., and we define the rescaled potential
\begin{equation}
\hat V(\hat k',\hat k,\hat p,\Lambda)  =  \frac{M\,\Lambda}{\pi^2}\,V(\Lambda
\hat k',\Lambda \hat k,\Lambda \hat p,\Lambda).
\label{potresc}
\end{equation}
This converts Eq.~(\ref{RGE1}) into the form of an RG equation,
\begin{equation}
\Lambda\, \frac{\partial \hat V}{\partial \Lambda}
 = \hat k'\,\frac{\partial \hat V}{\partial \hat k'}
+\hat k\,\frac{\partial \hat V}{\partial \hat k}
+\hat p\,\frac{\partial \hat V}{\partial \hat p}
 +  \hat V(\hat k',\hat k,\hat p,\Lambda) +\hat V(\hat k',1,\hat
p,\Lambda)\,\frac{1}{1-\hat p^2}\,\hat V(1,\hat k,\hat
p,\Lambda)\,.\label{RGE2}
\end{equation}
This has a similar structure to the RG equations that govern the
evolution of interactions in other areas of physics \cite{Wilson:1973jj}.
In this rescaled equation, the cutoff $\Lambda$, which can be
thought of as the highest acceptable low-energy scale, is the only
dimensioned quantity. The scaling of the potential with $\Lambda$ is directly
related to the dependence on the original low-energy variables, $p$, $k$ and $k'$,
as can be seen from the logarithmic derivatives on the right-hand side of
Eq.~(\ref{RGE2}).

As $\Lambda\rightarrow 0$, the solutions to Eq.~(\ref{RGE2}) tend to
fixed points, independent of $\Lambda$.\footnote{In some cases
the flows can drive the potential to infinity at a finite value of
$\Lambda$. In such cases it is better to follow the flow of the inverse of
the potential, which simply passes through zero and continues towards a
fixed point.}
This is because, for low enough values of $\Lambda$,
all memory of the scales of the underlying physics is lost and $\Lambda$ becomes
the only scale controlling the dependence of the potential on energy and momentum.
Expressed in units of $\Lambda$, the corresponding rescaled potential becomes
a constant.

These fixed points describe scale-invariant systems. For a system that lies close
to one of these points, we can use the RG flow near that point to define a systematic expansion of the potential in powers of the low-energy scales. The resulting
power counting can be used to organise the terms in a low-energy effective theory.
In Ref.~\cite{Birse:1998dk}, two fixed points were identified. One is just the
trivial point, $\hat V=0$. This is a stable point since all the perturbations
around it are irrelevant, that is, they flow towards it as $\Lambda\rightarrow 0$.
In fact, their scaling with $\Lambda$ follows from naive dimensional analysis
and the resulting expansion can be used to describe weakly interacting systems.

The second fixed point is the momentum-independent one that describes scattering
in the unitary limit. As described in Ref.~\cite{Birse:1998dk}, this is an
unstable point, with one relevant perturbation which corresponds to
the scattering length $a$. The expansion around this point can be used to describe
systems where the scattering length is much larger than the range of the forces.
In the power counting that controls this expansion, the terms are
promoted by two orders relative to naive dimensional analysis \cite{Birse:1998dk}.
The coefficients of these terms
are directly related to those of the effective-range expansion, and the counting
reflects the enhancement of the corresponding terms in scattering amplitude
by a factor of $1/a^2$.

Potentials corresponding to fine-tuned systems with $1/a=0$ lie on a critical surface
\cite{Wilson:1973jj,Morris:1998da} and flow towards
the unitary fixed point as $\Lambda\rightarrow 0$. If $1/a$ is nonzero, then the relevant
perturbation drives the flow away from the unitary fixed point for
$\Lambda\lesssim 1/a$ and towards the trivial point as $\Lambda\rightarrow 0$.
This reflects the fact that at very low energies, scattering can be treated
perturbatively, at least so long as the scattering length is finite.
The RG flow line linking the two fixed points is known as a
``renormalised trajectory" \cite{Wilson:1973jj,Morris:1998da}. The theory corresponding to this trajectory is renormalisable, both perturbatively
and nonperturbatively, in terms of a single coupling constant 
(in this case, the coefficient of the energy- and momentum-independent contact interaction).

\section{More fixed points}

To find further fixed points of the two-body system, we consider
here potentials that can be expressed as bivariate polynomials in the off-shell
momenta $k$ and $k'$, whose coefficients are functions of the on-shell
energy $p^2$. The structure of the nonlinear term in the RG equation
(\ref{RGE2}) means that no approximation is involved in choosing this
\textit{ansatz}. It allows us not only to identify
the fixed points but also to follow the RG flow lines in their vicinity. These
include the renormalised trajectories which run from one fixed point
to another.

If we restrict our potentials to be Hermitian, as these are of most interest,
then our \textit{ansatz} for them is conveniently written in a separable
matrix form,
\begin{equation}
V(k',k,p,\Lambda)= \chi^{\scriptscriptstyle\rm T}(k')\,\omega(p,\Lambda)\, \chi(k),
\label{Vsep}
\end{equation}
where $\omega$ is an $N\times N$, matrix. Here $\chi(k)$ is
defined as a column vector of powers of momentum and
$\chi^{\scriptscriptstyle\rm T}(k)$ is its transpose,
\begin{equation}
\chi^{\scriptscriptstyle\rm T}(k)=\bigl(k^{2n_1},\dots,k^{2n_N}\bigr),
\end{equation}
where $\bar S_N=\{n_1,\dots,n_N\}$ is a set of $N$ non-negative integers.
The $K$ matrix for this potential has a similar separable form,
\begin{equation}
K(k',k,p)= \chi^{\scriptscriptstyle\rm T}(k')\,\kappa(p)\, \chi(k).
\label{Tsep}
\end{equation}

From the Lippmann-Schwinger equation (\ref{LSEquationS}), we find that
$\kappa(p)$ can be related to $\omega(p,\Lambda)$ by
\begin{equation}
\kappa(p)^{-1}=\omega(p,\Lambda)^{-1}-\mathcal{G}(p,\Lambda),
\label{LSEkappa}
\end{equation}
where
$\mathcal{G}(p,\Lambda)$ is the matrix
\begin{equation}
\mathcal{G}(p,\Lambda) =  2M\,\mathcal{P}\! \int \frac{d^3l\,\theta(\Lambda-l)}{(2
\pi)^3}\,\frac{\chi(l)\,\chi^{\scriptscriptstyle\rm T}(l)}{p^2-l^2}\, .
\label{gmatS}
\end{equation}
The elements of this can be written as
\begin{equation}
\mathcal{G}_{ij}(p,\Lambda) = \frac{M}{\pi^2}\,I_{n_i+n_j}(p,\Lambda),
\end{equation}
where the regularised loop integrals are
\begin{equation}
I_n(p,\Lambda)=-\sum_{m=0}^n\frac{\Lambda^{2m+1}p^{2(n-m)}}{2m+1}
+\frac{p^{2n+1}}{2}\ln\frac{\Lambda+p}{\Lambda-p}\, .
\label{loopint}
\end{equation}

The rescaled version of the potential (\ref{Vsep}) can be written analogously as
\begin{equation}
\hat V(\hat k',\hat k,\hat p,\Lambda)= \chi^{\scriptscriptstyle\rm T}(\hat k')
\,\hat\omega(\hat p,\Lambda)\,
\chi(\hat k),
\label{rescVsep}
\end{equation}
where the elements of $\hat\omega(\hat p,\Lambda)$ are
\begin{equation}
\hat\omega_{ij}(\hat p,\Lambda)=\frac{M}{\pi^2}\,\Lambda^{2(n_i+n_j)+1}
\omega_{ij}(\Lambda\hat p,\Lambda).
\label{rescom}
\end{equation}
Inserting this into the RG equation (\ref{RGE2}), we find that
the evolution of $\hat\omega(\hat p,\Lambda)$ with $\Lambda$
is governed by the equation
\begin{equation}
\Lambda\, \frac{\partial \hat \omega}{\partial \Lambda}
=\hat p\,\frac{\partial \hat\omega}{\partial \hat p}
+2\,D(S_N)\,\hat\omega+2\,\hat\omega\, D(S_N)+\hat\omega
+\hat\omega\,\frac{\chi(1)\chi(1)^{\scriptscriptstyle\rm T}}{1-\hat p^2}
\,\hat\omega,
\label{RGEsep1}
\end{equation}
where $D(S_N)$ is the diagonal matrix of the elements of $S_N$.

This equation can be more easily solved if, following Ref.~\cite{Barford:2002je},
it is rewritten as a linear equation for $\hat\omega^{-1}$,\footnote{Further fixed points are obtained by considering non-invertible $\omega$ matrices. We do not deal with that case here.}
\begin{equation}
\Lambda\, \frac{\partial \hat \omega^{-1}}{\partial \Lambda}
=\hat p\,\frac{\partial \hat\omega^{-1}}{\partial \hat p}
-2\,\hat\omega^{-1} D(S_N)-2\,D(S_N)\,\hat\omega^{-1}
-\hat\omega^{-1}
-\frac{\chi(1)\chi(1)^{\scriptscriptstyle\rm T}}{1-\hat p^2}\,.
\label{RGEsep2}
\end{equation}
In this form, we can see that each of the elements of $\hat\omega^{-1}$
satisfies an uncoupled RG equation,
\begin{equation}
\Lambda\,\frac{\partial \left[\hat\omega^{-1}\right]_{ij}}{\partial \Lambda}
=\hat p\,\frac{\partial \left[\hat\omega^{-1}\right]_{ij}}{\partial \hat p}
-(2n_i+2n_j+1)\left[\hat\omega^{-1}\right]_{ij}-\frac{1}{1-\hat p^2},
\label{RGEsep2el}
\end{equation}
which can be integrated straightforwardly.

For any set of numbers $S_N$, we can find a nontrivial fixed-point solution,
$\omega_0(\hat p)$, to Eq.~(\ref{RGEsep2}), whose elements satisfy the ODEs
\begin{equation}
\hat p\,\frac{\partial \left[\hat\omega_0^{-1}\right]_{ij}}{\partial \hat p}
=(2n_i+2n_j+1)\left[\hat\omega_0^{-1}\right]_{ij}+\frac{1}{1-\hat p^2}.
\end{equation}
{This should satisfy the boundary condition that the matrix $\omega$ be analytic in $\hat p^2$
as $\hat p\rightarrow 0$ (or, in other words, it should be analytic in the
energy). Taking into account Eq.~(\ref{LSEkappa}) we obtain for the elements of the resulting matrix}
\begin{equation}
\left[\hat\omega_0^{-1}\right]_{ij}=C_{ij}\,\hat p^{2n_i+2n_j+1}+\hat I_{n_i+n_j}(\hat p),
\end{equation}
where $C_{ij}$ are arbitrary and we have introduced rescaled versions of the loop integrals (\ref{loopint}),
\begin{equation}
\hat I_n(\hat p)=\frac{1}{\Lambda^{2n+1}}\,I_n(\Lambda\hat p,\Lambda)
=-\sum_{m=0}^n\frac{\hat p^{2(n-m)}}{2m+1}
+\frac{\hat p^{2n+1}}{2}\ln\frac{1+\hat p}{1-\hat p}\,.
\end{equation}
Demanding analyticity of the potential in $p^2$ at $p^2=0$ means that we have to take $C_{ij}=0$.
When we undo the rescaling of Eq.~(\ref{rescom}), we find that, in physical units,
\begin{equation}
\omega_0^{-1}(p,\Lambda)=\mathcal{G}(p,\Lambda).
\label{FPom}
\end{equation}

From Eqs.~(\ref{LSEkappa}) and (\ref{FPom}) we see that each fixed point corresponds to a $K$
matrix with vanishing $\kappa(p)^{-1}$ or, provided $n_1=0$, infinite scattering
length. For example, the case $N=1$ and $S_1=\{0\}$ gives the unitary fixed point
as in Ref.~\cite{Birse:1998dk},
\begin{equation}
\hat V_{U}(\hat p)=\frac{1}{\hat I_0(\hat p)}
=-\left[1-\frac{\hat p}{2}\ln\frac{1+\hat p}{1-\hat p}\right]^{-1}.
\end{equation}
Adding an energy-independent perturbation to $\left[\hat\omega_0^{-1}\right]_{11}$
leads to a solution to Eq.~(\ref{RGEsep2el}) with the form
\begin{equation}
\left[\hat\omega^{-1}\right]_{11}=\frac{\alpha}{\Lambda}+\hat I_{0}(\hat p).
\end{equation}
The corresponding $K$ matrix is
\begin{equation}
{\cal K}(p)=\frac{\pi^2}{M\alpha},
\end{equation}
and so the parameter $\alpha$ is related to the physical scattering length by
\begin{equation}
\alpha=\frac{\pi}{2 a}.
\end{equation}
This perturbation grows as $\Lambda$ is lowered and so it is a relevant one.
It defines a renormalised trajectory which consists of the potentials
\begin{equation}
\hat V(\hat p,\Lambda)=\left[\frac{\alpha}{\Lambda}
-1+\frac{\hat p}{2}\ln\frac{1+\hat p}{1-\hat p}\right]^{-1}.
\label{RT1}
\end{equation}
This flows from unitary fixed point for $\Lambda\gg 1/a$ to the trivial one as
$\Lambda\rightarrow 0$. All other perturbations around the unitary point are
irrelevant \cite{Birse:1998dk} and so for $\alpha=0$ the other perturbations define the critical
surface of potentials that flow into this point as $\Lambda\rightarrow 0$.

The more general \textit{ansatz} described above allows us to construct
an infinite number of other fixed points using different sets $S_N$ of powers
of the off-shell momenta. The simplest of these points have a one-term-separable
structure. For example, the one with $N=1$ and $S_1=\{1\}$ is
\begin{equation}
\hat V_{S}(\hat k',\hat k,\hat p)
=\frac{\hat k^{\prime 2}\hat k^2}{\hat I_2(\hat p)}
=-\hat k^{\prime 2}\left[\frac{1}{5}+\frac{\hat p^2}{3}+\hat p^4
-\frac{\hat p^5}{2}\ln\frac{1+\hat p}{1-\hat p}\right]^{-1}\hat k^2.
\end{equation}
The solution to the RG equation (\ref{RGEsep2}),
\begin{equation}
\hat V(\hat k',\hat k,\hat p,\Lambda)
=\hat k^{\prime 2}\left[\frac{\gamma}{\Lambda^5}
+\frac{\eta\,\hat p^2}{\Lambda^3}
+\frac{\sigma\,\hat p^4}{\Lambda}+\hat I_2(\hat p)\right]^{-1}
\hat k^2,
\label{RT2}
\end{equation}
shows that this fixed point has three relevant perturbations. This means that
it would describe triply fine-tuned systems, which makes it very unlikely to
be realised in practice. The potentials (\ref{RT2}) that contain only these
perturbations form a three-parameter family of renormalised trajectories that
run from $\hat V_{S}$ to the trivial point.

At the same order in the off-shell momenta, there is also a fixed point with
$N=2$ and $S_2=\{0,1\}$:
\begin{equation}
\hat V_{L}(\hat k',\hat k,\hat p)=\left(1,\;\hat k^{\prime 2}\right)
\left(\begin{array}{cc}\hat I_0(\hat p)&\hat I_1(\hat p)\cr\noalign{\vspace{5pt}}
\hat I_1(\hat p)&\hat I_2(\hat p)\end{array}\right)^{-1}
\left(\begin{array}{c}1\cr\noalign{\vspace{5pt}}\hat k^2\end{array}\right).
\end{equation}
This is the simplest example of a fixed point that does not have
a one-term separable form. The existence of such a point had been hinted at
previously \cite{cohen,Harada:2006cw} but no explicit expression for it was
found. It has six relevant perturbations, which can be seen
in the potential,
\begin{equation}
\hat V(\hat k',\hat k,\hat p,\Lambda)=\left(1,\;\hat k^{\prime 2}\right)
\left(\begin{array}{cc}{\displaystyle\frac{\alpha}{\Lambda}+\hat I_0(\hat p)}
&{\displaystyle\frac{\beta}{\Lambda^3}+\frac{\delta\,\hat p^2}{\Lambda}
+\hat I_1(\hat p)}\cr\noalign{\vspace{5pt}}
{\displaystyle\frac{\beta}{\Lambda^3}+\frac{\delta\,\hat p^2}{\Lambda}
+\hat I_1(\hat p)}
&{\displaystyle\frac{\gamma}{\Lambda^5}+\frac{\zeta\,\hat p^2}{\Lambda^3}
+\frac{\eta\,\hat p^4}{\Lambda}+\hat I_2(\hat p)}\end{array}\right)^{-1}
\left(\begin{array}{c}1\cr\noalign{\vspace{5pt}}\hat k^2\end{array}\right),
\label{RT3}
\end{equation}
which satisfies the RG equation (\ref{RGEsep2}). The on-shell $K$ matrix for
this is
\begin{equation}
{\cal K}(p)=\frac{\pi^2}{M}\,\frac{\gamma+\zeta\,p^2+\eta\,p^4
-2(\beta+\delta\, p^2)p^2+\alpha\,p^4}{\alpha(\gamma+\zeta\,p^2+\eta\,p^4)
-(\beta+\delta\, p^2)^2}\, .
\end{equation}
The fixed point describes the scale-free limit where all of the parameters
$\alpha$, \dots, $\eta$ vanish. However the corresponding scattering amplitude
is not uniquely defined until one specifies how this limit is taken.

The renormalised trajectories that flow out of this fixed point can be followed
more easily by rewriting Eq.~(\ref{RT3}) in the form
\begin{equation}
\hat V(\hat k',\hat k,\hat p,\Lambda)
=\left(1,\;\hat k^{\prime 2}\right)\det\left[\omega(\hat p,\Lambda)\right]
\left(\begin{array}{cc}{\displaystyle\frac{\gamma}{\Lambda^5}
+\frac{\zeta\,\hat p^2}{\Lambda^3}+\frac{\eta\,\hat p^4}{\Lambda}
+\hat I_2(\hat p)}
&{\displaystyle-\,\frac{\beta}{\Lambda^3}-\frac{\delta\,\hat p^2}{\Lambda}
-\hat I_1(\hat p)}\cr\noalign{\vspace{5pt}}
{\displaystyle-\,\frac{\beta}{\Lambda^3}-\frac{\delta\,\hat p^2}{\Lambda}
-\hat I_1(\hat p)}
&{\displaystyle\frac{\alpha}{\Lambda}+\hat I_0(\hat p)}\end{array}\right)
\left(\begin{array}{c}1\cr\noalign{\vspace{5pt}}\hat k^2\end{array}\right),
\label{RT3a}
\end{equation}
where
\begin{equation}
\det\left[\omega(\hat p,\Lambda)\right]
=\left[\left(\frac{\alpha}{\Lambda}+\hat I_0(\hat p)\right)
\left(\frac{\gamma}{\Lambda^5}+\frac{\zeta\,\hat p^2}{\Lambda^3}
+\frac{\eta\,\hat p^4}{\Lambda}+\hat I_2(\hat p)\right)
-\left(\frac{\beta}{\Lambda^3}+\frac{\delta\,\hat p^2}{\Lambda}
+\hat I_1(\hat p)\right)^2\right]^{-1}.
\end{equation}
In general these potentials run to the trivial fixed point. For example, in the
case that $\alpha\gamma-\beta^2$ is nonzero, the determinant behaves for small
$\Lambda$ as
\begin{equation}
\det\left[\omega(\hat p)\right]=\frac{\Lambda^6}{\alpha\gamma-\beta^2}
+{\cal O}(\Lambda^7).
\end{equation}
As a result, all elements of the potential (\ref{RT3a}) vanish at least
linearly in $\Lambda$ as $\Lambda\rightarrow 0$.

In the more fine-tuned case where $\gamma$ is nonzero but $\alpha\gamma-\beta^2=0$,
the determinant behaves as
\begin{equation}
\det\left[\omega(\hat p)\right]=\frac{\Lambda^5}{\gamma \hat I_0(\hat p)}
+{\cal O}(\Lambda^6),
\end{equation}
and all elements of the potential vanish except for $\hat V_{11}$. For small
$\Lambda$ this has the form
\begin{equation}
\hat V_{11}(\hat p,\Lambda)=\frac{1}{\hat I_0(\hat p)}+{\cal O}(\Lambda),
\end{equation}
which is just the unitary fixed point in the limit $\Lambda\rightarrow 0$.
Finally, in the case that the only nonzero relevant perturbation is $\alpha$,
we get a potential that runs to the separable fixed point $\hat V_S$.

The other perturbations around each of these fixed points are all irrelevant. They
involve either higher powers of the energy ($p^2$) or different powers of the
off-shell momenta. The scaling of the former can be found easily by adding
additional energy-dependent terms to the potentials just discussed. For the latter,
we can use a more general version of Eq.(\ref{RGEsep1}) that contains all powers of
$k^{\prime 2}$ and $k^2$, not just the ones that appear in the fixed point.
Adding these perturbations leads to a critical surface for each point,
consisting of all potentials that flow to that point as $\Lambda\rightarrow 0$.

In cases such as $S$-wave nucleon-nucleon scattering where the coefficients of the relevant perturbations are unnaturally small, the potential lies close to the
critical surface for large cutoffs. As $\Lambda$ is lowered the potential
initially runs towards the fixed point. Then, when $\Lambda$ becomes comparable
to the scales of the relevant perturbations (such as $1/a$), the flow
deviates from the critical surface and heads towards a renormalised trajectory
that leads to a different fixed point, generally the trivial one.
(See, for example, Fig.~1 of Ref.~\cite{Birse:1998dk}.)

\section{Summary}

In this paper we have outlined a separable matrix \textit{ansatz} for the
potentials that arise in EFT descriptions of two-body systems with short-range
interactions. This provides a tool for constructing new fixed points of
the RG for these systems, as well as the renormalised trajectories connecting
them. In particular we are able to construct a fixed point whose existence
has previously been only conjectured.
These new fixed points indicate a much
richer structure than previously recognized in the RG flows of simple
short-range potentials. Each of them is unstable in at least two directions and so a physical system described by 
one of them would need to have fine-tuned values for at least two parameters.

\acknowledgments

This work was supported in part by the Georgian Shota Rustaveli National
Science Foundation (grant FR/417/6-100/14), by the DFG (grant SFB/TR 16,
``Subnuclear Structure of Matter''), by the UK STFC (grant ST/J000159/1),
by the EC Research Infrastructure Integrating Activity ``Study of
Strongly Interacting Matter'' (grant agreement no.~283286),
and by the ERC (project 259218 NUCLEAREFT). The authors are grateful to the
Centro de Ciencias de Benasque Pedro Pascual for its hospitality, and to the
organisers of the workshop on ``Bound states and resonances in EFTs and
lattice QCD".



\begin{thebibliography}{99}

\bibitem{Wilson:1973jj}
  K.~G.~Wilson and J.~B.~Kogut,
  Phys.\ Rept.\  {\bf 12}, 75 (1974).

\bibitem{Morris:1998da}
  T.~R.~Morris,
  Prog.\ Theor.\ Phys.\ Suppl.\  {\bf 131}, 395 (1998).

\bibitem{Meurice11}
``New applications of the renormalization group in nuclear, particle and condensed matter physics,'' edited by Y. Meurice, R. Perry and S.-W. Tsai,
Phil. Trans. Roy. Soc. Lond. A {\bf 369} (2011).

%
\bibitem{Weinberg:um}
S.~Weinberg,
Nucl.\ Phys.\ {\bf B363}, 3 (1991).

\bibitem{Gegelia:gn}
J.~Gegelia,
Phys.\ Lett.\ B {\bf 429}, 227 (1998).

\bibitem{Kaplan:1998tg}
  D.~B.~Kaplan, M.~J.~Savage and M.~B.~Wise,
  Phys.\ Lett.\ B {\bf 424}, 390 (1998).

\bibitem{vanKolck:1998bw}
  U.~van Kolck,
  Nucl.\ Phys.\ A {\bf 645}, 273 (1999).

\bibitem{Birse:1998dk}
  M.~C.~Birse, J.~A.~McGovern and K.~G.~Richardson,
  Phys.\ Lett.\  B {\bf 464}, 169 (1999).

\bibitem{Bethe:1949yr}
  H.~A.~Bethe,
  Phys.\ Rev.\  {\bf 76}, 38 (1949).

\bibitem{Birse:2010fj}
  M.~C.~Birse,
  Phil.\ Trans.\ Roy.\ Soc.\ Lond.\ A {\bf 369}, 2662 (2011).

\bibitem{cohen} T. D. Cohen, private communication.

\bibitem{Harada:2006cw}
  K.~Harada and H.~Kubo,
  Nucl.\ Phys.\ B {\bf 758}, 304 (2006).

\bibitem{Bogner:2003wn}
  S.~K.~Bogner, T.~T.~S.~Kuo and A.~Schwenk,
  Phys.\ Rept.\  {\bf 386}, 1 (2003).

\bibitem{Barford:2002je}
  T.~Barford and M.~C.~Birse,
  Phys.\ Rev.\ C {\bf 67}, 064006 (2003).

\end{thebibliography}
\end{document}